\documentclass[12pt]{article}

 \usepackage{paralist} 
     \usepackage{placeins} 
     \usepackage{amsmath,amsthm,amssymb}
     \usepackage{epsfig}
     \usepackage{psfrag}
     \usepackage{tabls}
     \usepackage{supertabular}
     \usepackage[square]{natbib}
     \renewcommand{\cite}{\citep}
     \usepackage[top=2.5cm, bottom=2.5cm, left=2.5cm, right=2.5cm]{geometry}
\usepackage{indentfirst} 
\usepackage{colortbl} 
\definecolor{niceyellow}{rgb}{0.98,0.92,0.73}
\usepackage{fancyhdr}
\usepackage{fancybox}
\usepackage{hyperref} 
    \newcommand{\myfecha}{\today}

\begin{document}
\begin{center}
\Large{\bf  
{\em Physics Education Research} and the Teaching and Learning of Physics 
}
\end{center}
\begin{flushleft}
{\bf Sergio Rojas}\\
srojas@usb.ve \\
{Physics Department, Universidad Sim\'{o}n Bol\'{\i}var,
Ofic. 220, Apdo. 89000, Caracas 1080A, Venezuela.}
\end{flushleft}

{\bf Abstract}.
A brief account of some recent controversies about the teaching
and learning of physics is presented. A shorter version of this outcome
was accepted by \emph{The Physics Teacher}, but publication is still pending.

\bigskip

{\bf Keywords}: Physics Education Research; Students Performance; Mathematics
and Physics.
\setlength{\parindent}{0.75cm}

\bigskip

In the October 2009 issue of \emph{The Physics Teacher}, we
were delighted to read a lively and enlightening genuine dialog on the
important process of teaching and learning
physics.\cite{Sobel:2009a,Lasry:2009a,Sobel:2009b,Lasry:2009b}
The lively discussion reminded me of recently
debates\cite{Glazek:2008,Klein:2007,KleinDebate1:2007,KleinDebate2:2007,KleinDebate3:2007} 
which bring to light some key issues regarding the teaching and
learning of physics which are dissected in published literature and
in university hallways, reflected in statements such as ``I have
always been skeptical of general methods, tools, and jargon emerging
from the inexact science, research into teaching, even though it has
produced some interesting
results.''\cite{Vogt:2007}  

Professor Wieman has also called for
cautiousness when measuring teaching outcomes as one could create
illusions about what students actually
learn\cite{Wieman:2007},
or in the words of Professor Sobel ``Yes, in a special (possibly
grant{}-supported) program, with smaller groups, with highly motivated
instructors and students, with less content, students might do well,
but that's not the real
world.''\cite{Sobel:2009b}

One could argue that these opinions result
from the observation that physics is an intrinsically quantitative
based subject, and from the belief that it is in physics classes where
students should actually get training to apply what they have learned
in their math classes. Yet, much of the recent
{\textit{Physics Education Research}}
seems to overemphasize the
importance of teaching the qualitative or conceptual physical
aspects,\cite{MualemEylon:2007,HoellwarthEtAll:2005,SabellaRedish:2007,Walsh:2007}
and to deemphasize the significance of standard mathematical
reasoning, which are crucial for understanding physical processes, and
which are not stressed, or even taught, because, rephrasing a passage
from a recent editorial,\cite{Klein:2007}
they interfere with the students' emerging sense of physical insight.
In addition, the lack of sensitivity to professors who want to be
better teachers and to students who want to do well in their physics
studies is further demonstrated in a letter written by a physics
teacher to {\textit{The Assessment and Qualifications Alliance}} (AQA)
and to the UK Department for
Education,\cite{TeacherBeg}
and by controversial outcomes coming from some publicized instructional
practices.\cite{Ates:2007,Coletta:2008,Ates:2008}

Consequently, physics instructors continue to
face the problem of finding suitable advice on how to approach the
teaching of physics in the most efficient way and an answer to the
question of how much time should be spent on intuitive, conceptual
reasoning and how much time in developing 
quantitative reasoning \cite{Rojas:2008,Rojas:2009ajp,Rojas:2010rmf}.

According to Professor Reif, perhaps the
answer to this quandary is that ``In science education the primary
interest is not focused on the science itself, but on students who are
trying to learn scientific knowledge and thinking. A truly scientific
approach to education would thus need to strive for a better
understanding of the underlying human thought processes and knowledge
required for good performance in particular scientific
domains.''\cite{Reif:2008}
After outlining some basic issues on how to approach the question of
performance in complex domains, Professor Reif further reminds us about
the delicate complexity of the problem of teaching and learning by
telling the story on how he was able to pronounce the sound
corresponding to the letter ``r'' in
English.\cite{Reif:2008a}



\label{LastPage}

\end{document}